\documentclass[aps,preprint,prd,showpacs,superscriptaddress,amssymb,amsmath,preprintnumbers,floatfix]{revtex4-1}
\usepackage{graphicx}
\usepackage{subfigure}
\usepackage{hyperref}
\allowdisplaybreaks[1]

\newcommand{\e}{{\rm{e}}}
\newcommand{\img}{{\rm{i}}}

\def\slashchar#1{\setbox0=\hbox{$#1$} 
\dimen0=\wd0 
\setbox1=\hbox{/} \dimen1=\wd1 
\ifdim\dimen0>\dimen1 
\rlap{\hbox to \dimen0{\hfil/\hfil}} 
#1 
\else 
\rlap{\hbox to \dimen1{\hfil$#1$\hfil}} 
/ 
\fi}

\begin{document}
\title{
Unphysical poles and dispersion relations for M\"obius domain-wall fermions in free field theory at finite $L_s$
} 

\newcommand{\Columbia}{
Physics Department,
Columbia University,
New York 10027, USA
}

\author{Masaaki~Tomii}
\email{mt3164_at_columbia.edu}
\affiliation{\Columbia}


\begin{abstract}
We find that the quark propagator constructed from the domain-wall fermion
operator has $L_s-1$ extra poles as well as the pole that realizes the
physical quark in the continuum limit.
We show the energy-momentum dispersion relation for the physical and
unphysical poles of M\"obius domain-wall fermions in free field theory at
finite $L_s$.
The dependence of extra pole energies on the M\"obius parameter $b-c$
and on the domain-wall height $M_5$ is investigated.
Our result suggests that small values of $b-c$ set a large lower bound on
the unphysical pole masses and the contribution of these poles could be
suppressed well by calculating with small $b-c$.
\end{abstract}

\maketitle

\section{Introduction}
\label{sec:intro}

Introducing the charm quark into lattice simulation is desired to
provide accurate Standard Model predictions for flavor physics, which
enable us to probe for new physics beyond the Standard Model.
Especially, the non-perturbative calculation of quantities associated with
the Glashow-Iliopoulos-Maiani (GIM) mechanism \cite{Glashow:1970gm}
such as the $K_L$--$K_S$ mass difference ($\Delta M_K$) essentially needs
the charm quark to cancel the divergent contributions of up-quark loop diagrams.
Because of the large charm-quark mass $m_c$ compared to the typical scale
of QCD, a lattice calculation including a charm quark encounters a scale problem.
Namely, the lattice cutoff $a^{-1}$ needs to be sufficiently larger than $m_c$
to safely control the discretization error arising from $am_c$, while the box size
$L$ is usually required to obey $m_\pi L\gtrsim4$, with the pion mass $m_\pi$,
to avoid uncontrollable finite volume effects.
Thus, lattice calculation at the physical pion and charm quark masses
is a challenging task for the currently available computational resources.

This work is devoted to investigating properties of the discretization effects
appearing in M\"obius domain-wall fermions \cite{Brower:2004xi,Brower:2012vk},
an extension of Shamir domain-wall fermions \cite{Kaplan:1992bt,Shamir:1993zy},
at heavy quark masses.
Although the charm quark completely violates chiral symmetry due to
its heavy mass, introducing the charm quark as a domain-wall fermion
is still necessary to achieve an accurate GIM cancellation if the light quarks
are implemented with a domain-wall fermion formulation, which appropriately
preserves the chiral symmetry of the light quarks.
There have been several works on $D$ meson decay constants using
domain-wall fermions \cite{Chen:2014hva,Boyle:2017jwu} and overlap
fermions for valence quarks and domain-wall fermions for sea quarks
\cite{Yang:2014sea}.
A lattice simulation including $2+1+1$ optimal domain-wall fermions \cite{Chiu:2002ir}
was implemented \cite{Chen:2017kxr} and was the first study with a dynamical
domain-wall charm quark.
In addition, the RBC and UKQCD collaborations are pursuing the calculation
of the $\Delta M_K$ \cite{Bai:2014cva,Christ:2014qwa},
$\varepsilon_K$ \cite{Christ:2014qwa,Bai:2016gzv}, rare kaon decays
$K\rightarrow\pi l^+l^-$ \cite{Christ:2015aha,Christ:2016mmq} and
$K\rightarrow\pi\nu\bar\nu$ \cite{Christ:2016eae,Bai:2017fkh}, which are
all associated with the GIM mechanism and quite sensitive to the discretization
effects due to the charm-quark mass.

The charm quark treated in the domain-wall fermion formulation is supposed
to have some special difficulties in addition to the na\"\i ve $O(a^2m^2_c)$
discretization errors and beyond.
The seminal work on domain-wall fermions at large quark masses 
\cite{Liu:2003kp,Christ:2004gc} investigated the hermitian version of the domain-wall
operator, the five-dimensional Dirac operator multiplied by the chirality operator
$\gamma_5$ and the five-dimensional reflection operator.
It found that the hermitian operator contains unphysical modes whose
eigenvalues are largely independent of the input quark mass.
This fact indicates that as the input quark mass approaches the cutoff
the contribution of physical modes will be contaminated by unphysical
modes.

This unphysical contribution may be related to the oscillatory behavior of
domain-wall fermions
\cite{Dudek:2006ej}, which is a particular issue of domain-wall fermions
and is observed in correlation functions when a simulation is carried out at large
domain-wall heights such as $M_5\simeq1.7$.
This unphysical oscillation was understood as the result of negative eigenvalues
of the transfer matrix \cite{Syritsyn:2007mp}, which were shown to exist in
the region of $M_5>1$ in the free field case.

Recently, another description of the origin of the unphysical oscillation was
proposed \cite{Liang:2013eoa,Sufian:2016cft}.
They argued that the four-dimensional quark propagator constructed from the
domain-wall fermion operator has an extra pole, whose energy has a non-zero
imaginary part $\img\pi$ in lattice units for $M_5>1$ in free field theory,
leading to an oscillatory behavior of the quark propagator.
Their numerical result \cite{Sufian:2016cft} indicates that the impact of the
unphysical oscillation could be reduced by choosing $M_5$ and the M\"obius
parameters $b$ and $c$ to satisfy $M_5(b-c)<1$ and that the Bori\c{c}i
domain-wall fermion \cite{Borici:1999zw} ($b=c$) is optimal to
suppress the unphysical oscillation.
Although this viewpoint of an unphysical pole is quite impressive and
provides a clear interpretation of the unphysical effects of domain-wall fermions,
we find that when examined in detail their description
\cite{Liang:2013eoa,Sufian:2016cft} is not correct.
Especially, we find there are $L_s-1$ unphysical poles, while they found
only one.

To motivate the presence of this collection of unphysical poles, it may be helpful to
consider the case of overlap fermions, which has the same quark propagator as
domain-wall fermions in the limit of infinite $L_s$ up to a contact term and a
normalization factor.
The overlap Dirac operator is given by
\begin{equation}
D_{\rm ov} = \frac{1}{2}+\frac{1}{2}\gamma_5 \frac{H}{\sqrt{H^2}},
\end{equation}
with a hermitian kernel $H$.
The corresponding propagator contains $\sqrt{H^2}$ and becomes ambiguous in
the region where $H^2$ is not positive definite, resulting in the presence of an
unphysical brunch cut for an imaginary value of Euclidean momentum variable
$p_4$.
In this paper, we demonstrate the presence of $L_s-1$ unphysical poles for
finite $L_s$ as the counterpart of this brunch cut.

We also examine the fundamental properties of the unphysical poles of
domain-wall fermions by showing the dispersion relation for the physical and
unphysical poles in free field theory at finite $L_s$.
We find that the range of unphysical pole energies significantly
depends on $M_5$ and $b-c$ as well as on the spatial momentum and that
small values of $b-c$ set a large lower bound on the unphysical pole energies,
possibly suppressing the contribution of the unphysical poles.
Since this range of unphysical pole energies is found to be independent of the
physical quark mass, numerical calculation with heavy quarks would be contaminated
by unphysical poles as the physical quark mass approaches the lower bound
of the unphysical pole region.

The paper is organized as follows.
In Section~\ref{sec:action}, we give definitions and some comments on the
parameters of M\"obius domain-wall fermions.
In Section~\ref{sec:qprop}, we give the five- and four-dimensional propagators
of M\"obius domain-wall fermions.
In Section~\ref{sec:unphys_poles}, we show the presence of unphysical poles
of domain-wall fermions as well as the physical pole.
In Section~\ref{sec:disp_rel}, we show the dispersion relation for the physical and
unphysical poles and discuss the dependence of unphysical pole energies on
the parameters of M\"obius domain-wall fermions.
In Section~\ref{sec:conclusion}, we conclude this paper and give some discussion.
In Appendix~\ref{sec:infLs}, we discuss a connection with overlap fermions and
demonstrate that domain-wall fermions in the limit of infinite $L_s$ and overlap
fermions have an unphysical branch cut instead of unphysical poles.
In Appendix~\ref{sec:qprop_special}, we give the five- and four-dimensional
propagators in some special cases, in which the usual form of these propagators
is irrelevant.

\section{M\"obius domain-wall fermions}
\label{sec:action}

In this study, we work in the momentum space, where the lattice
action of a M\"obius domain-wall fermion is given by
\begin{equation}
S = \sum_{s,t=0}^{L_s-1}\sum_p\overline\psi_s(-p) (D_{\rm MDW})_{s,t}\psi_t(p).
\end{equation}
Here, $\overline\psi_s(-p)$ and $\psi_t(p)$ are the five-dimensional M\"obius
domain-wall fermion fields labeled by the four-dimensional momentum
variables, $-p$ and $p$, and the indices for the fifth direction, $s,t = 0,1,\ldots,L_s-1$.
We employ the convention used by the RBC and UKQCD collaborations
\cite{Blum:2014tka}, in which the corresponding Dirac operator is defined by
\begin{align}
D_{\rm MDW} &= \left(
\begin{array}{cccccc}
\tilde D & -P_- & 0 & \ldots & 0 & mP_+\\
-P_+ & \tilde D & -P_- & \ddots & 0 & 0\\
0 & -P_+ & \tilde D & \ddots & \ddots & \vdots \\
\vdots & \ddots & \ddots & \ddots & -P_- & 0 \\
0 & 0 & \ddots & -P_+ & \tilde D & -P_-\\
mP_- & 0 & \ldots & 0 & -P_+ & \tilde D
\end{array}
\right).
\end{align}
Here, we define the chiral projection operators
$P_\pm = \frac{1}{2}(1\pm\gamma_5)$ and
\begin{equation}
\tilde D = D_-^{-1}D_+,\ \ 
D_+ = 1+bD_W,\ \ 
D_- = 1-cD_W,
\end{equation}
with the Wilson Dirac operator $D_W$ in the momentum space at a
negative mass parameter $-M_5$,
\begin{equation}
D_W = \img\slashchar{\tilde p}+\sum_\mu(1-\cos p_\mu) -M_5,
\end{equation}
where $\slashchar{\tilde p} = \sum_\mu\gamma_\mu\sin p_\mu$.
For simplicity, we omit the lattice spacing $a$ and everything is expressed
in lattice units throughout this paper.

We define the corresponding four-dimensional quark fields by
\begin{equation}
q = P_-\psi_0+P_+\psi_{L_s-1},\ \ 
\overline q = \overline\psi_0P_+ + \overline\psi_{L_s-1}P_-.
\label{eq:4Dquarks}
\end{equation}
As shown in \cite{Brower:2012vk,Blum:2014tka},
the four-dimensional quark propagator constructed from these fields
$S_F^{4d}(p) = \langle q(-p)\overline q(p)\rangle$ in the limit $L_s\rightarrow\infty$
is the same as that in the corresponding overlap action up to a contact term and
a normalization factor.

The action has five input parameters in total: the mass parameter $m$,
the extent of the fifth dimension $L_s$, the domain-wall height $M_5$,
the M\"obius parameters $b$ and $c$.
Except for the mass parameter $m$, these are parameters of the
regularization and do not affect any observables in the continuum limit.
Therefore we can tune them to minimize unwanted discretization effects.
As is well known, the fifth-dimensional extent $L_s$ determines the amount
of violation of chiral symmetry on the lattice, which is usually quantified
by the residual mass $m_{\rm res}$ and vanishes in the limit
$L_s\rightarrow\infty$.
The domain-wall height $M_5$ determines the scale for the exponential
locality of the four-dimensional effective fermion field
\cite{Hernandez:1998et,Kikukawa:2000kd} and is also related to $m_{\rm  res}$
\cite{Aoki:2002vt}.
The optimal choice of $M_5$ is 1 for the case of the free field, while
that in the non-perturbative case has been studied by analyzing the
spectral flow on some representative configurations to minimize
$m_{\rm res}$
\cite{Aoki:2002vt,Antonio:2008zz,Hagler:2007xi,WalkerLoud:2008bp}.
The obtained best choice was $M_5=1.7$--$1.8$
depending on the detail of the lattice setup.
By applying link smearing, the residual mass may be better controlled
and the optimal choice of $M_5$ could be moved to 1 \cite{Cho:2015ffa}.
In addition, the $M_5$-dependence of the amount of discretization error for
the heavy-heavy decay constant was investigated, resulting in a
slightly smaller tuned value $M_5=1.6$ \cite{Boyle:2016imm}.
The M\"obius scale $b+c$, which is proportional to the M\"obius kernel,
has also been tuned to minimize $m_{\rm res}$, while the dependence on
$b-c$ has not been studied a lot.
In this work, we investigate how the significance of the unphysical modes
depends on these parameters including $b-c$.

\section{Quark propagator at finite $L_s$}
\label{sec:qprop}

The five-dimensional Dirac operator $D_{\rm MDW}$ can be rewritten as
\begin{equation}
D_{\rm MDW} = {b+c\over D_-^\dag D_-}\img\slashchar{\tilde p} + W^+P_- + W^-P_+,
\end{equation}
where
\begin{align}
W^\pm_{s,t} &= W\delta_{s,t}-\delta_{s\pm1,t}+m\delta_{s/t,L_s-1}\delta_{t/s,0},
\label{eq:Wpm_st}
\\
W &= {-bc(\tilde p^2+{\cal M}^2)+(b-c){\cal M}+1\over D_-^\dag D_-},
\\
{\cal M}&=\sum_\mu(1-\cos p_\mu) -M_5,
\\
D_-^\dag D_- &= c^2(\tilde p^2+{\cal M}^2)-2c{\cal M}+1,
\end{align}
with $\tilde p^2 = \sum_\mu\sin^2p_\mu$.
Thus, we can calculate the five-dimensional propagator of M\"obius
domain-wall fermions in the same way \cite{Shamir:1993zy}
as for Shamir domain-wall fermions.
We obtain
\begin{align}
D^{-1}_{\rm MDW}
&= \left[-{b+c\over D_-^\dag D_-}\img\slashchar{\tilde p}+W^-\right]G^-P_-
+ \left[-{b+c\over D_-^\dag D_-}\img\slashchar{\tilde p}+W^+\right]G^+P_+,
\\
G^\pm &= \left[\bigg({b+c\over D_-^\dag D_-}\bigg)^2\tilde p^2+W^\mp W^\pm\right]^{-1}
\equiv (Q^\pm)^{-1},
\notag\\
G^\pm_{s,t}&= A_0\e^{-\alpha|s-t|}+A_\pm\e^{\alpha(s+t-L_s+1)}+A_\mp\e^{-\alpha(s+t-L_s+1)}
+A_m\cosh[\alpha(s-t)],
\label{eq:G_comp}
\\
\cosh\alpha &=
\frac{\big({b+c\over D_-^\dag D_-}\big)^2\tilde p^2 + W^2 + 1}{2W},
\\
A_0 &= \frac{1}{2W\sinh\alpha},
\\
A_\pm &= \frac{A_0}{F_{L_s}}(1-m^2)(W-\e^{\mp\alpha}),
\label{eq:Apm}
\\
A_m &= \frac{A_0}{F_{L_s}}
\left[
4mW\sinh\alpha-2(W\e^{-\alpha}-1+m^2(1-W\e^\alpha))\e^{-\alpha L_s}
\right],
\label{eq:Am}
\\
F_{L_s}
&= \e^{\alpha L_s}(1-W\e^\alpha+m^2(W\e^{-\alpha}-1))-4mW\sinh\alpha
\notag\\&\hspace{4mm}
+\e^{-\alpha L_s}(W\e^{-\alpha}-1+m^2(1-W\e^\alpha)).
\label{eq:FLs}
\end{align}

Since the four-dimensional effective fields $q$ and $\overline q$ are given by
\eqref{eq:4Dquarks}, the four-dimensional quark propagator
$S_F^{4d}(p) = \langle q(-p)\overline q(p)\rangle$ constructed from the M\"obius
domain-wall fermions is written as
\begin{align}
S_F^{4d}(p)
&= P_-(D_{\rm MDW}^{-1})_{0,0}P_+
+P_+(D_{\rm MDW}^{-1})_{L_s-1,L_s-1}P_-
\notag\\& \hspace{4mm}
+P_-(D_{\rm MDW}^{-1})_{0,L_s-1}P_-
+ P_+(D_{\rm MDW}^{-1})_{L_s-1,0}P_+
\notag\\
&= {2\sinh(\alpha L_s)\over F_{L_s}}{b+c\over D_-^\dag D_-}
\img\slashchar{\tilde p}
\notag\\&\hspace{4mm}
+{2\over F_{L_s}}
\left\{
m[W\sinh(\alpha(L_s-1))-\sinh(\alpha L_s)]
-W\sinh\alpha
\right\}.
\label{eq:qprop_4d}
\end{align}
In the limit of infinite $L_s$, this four-dimensional propagator becomes
\begin{equation}
S_F^{4d}(p) \xrightarrow{\ L_s\rightarrow\infty\ }
\frac{{b+c\over D_-^\dag D_-}\img\slashchar{\tilde p}+m(W\e^{-\alpha}-1)}
{1-W\e^\alpha+m^2(W\e^{-\alpha}-1)}. 
\label{eq:qprop_4d_infLs}
\end{equation}
Since the M\"obius scale $b+c$ is associated only with $L_s$-dependence,
the propagator in the limit of infinite $L_s$ \eqref{eq:qprop_4d_infLs} must
be independent of $b+c$, despite its apparent dependence on that combination.
The $b+c$-independent form is given in \eqref{eq:qprop_4d_infLs2}.

The four-dimensional quark propagator \eqref{eq:qprop_4d} at finite $L_s$ is
quite different from that given in \cite{Sufian:2016cft}.
This may originate from the slight difference in $F_{L_s}$.
Since the coefficients \eqref{eq:Apm} and \eqref{eq:Am} are determined through
the boundary conditions \eqref{eq:BC_init}--\eqref{eq:BC_last} for the fifth
direction, the validity of these coefficients and $F_{L_s}$ given in this section
can be checked by inserting $G^\pm_{s,t}$ into the boundary conditions.

\section{Physical and unphysical poles at finite $L_s$}
\label{sec:unphys_poles}

It is well known that the quark propagator $S_F^{4d}$ has a
physical pole that reproduces an appropriate Dirac fermion in the
continuum limit.
This can be verified by expanding the denominator of the quark propagator
with respect to the momentum variable.
In the case of infinite $L_s$, the denominator of $S_F^{4d}$ in
\eqref{eq:qprop_4d_infLs} is expanded as
\begin{equation}
1-W\e^\alpha+m^2(W\e^{-\alpha}-1)
= -\frac{(b + c)\big[M_5^2 (2 - (b - c) M_5)^2 m^2 + p^2\big]}
{M_5 (2 - (b - c) M_5) (1 + c M_5)^2} + O(p^4,m^2p^2).
\end{equation}
Thus, the mass $m_f^{\rm pole}$ of the physical pole is
approximated at $M_5(2-(b-c)M_5)m$ for a light quark and
is generally different from the input mass $m$ even in the
case of infinite $L_s$.
In this work, we input $m_f^{\rm pole}$ and tune the parameter
$m$ to realize the pole mass $m_f^{\rm pole}$.

\begin{figure}[t]
\begin{center}
\subfigure{\mbox{\raisebox{1mm}{\includegraphics[width=99mm, bb=0 0 345 230]{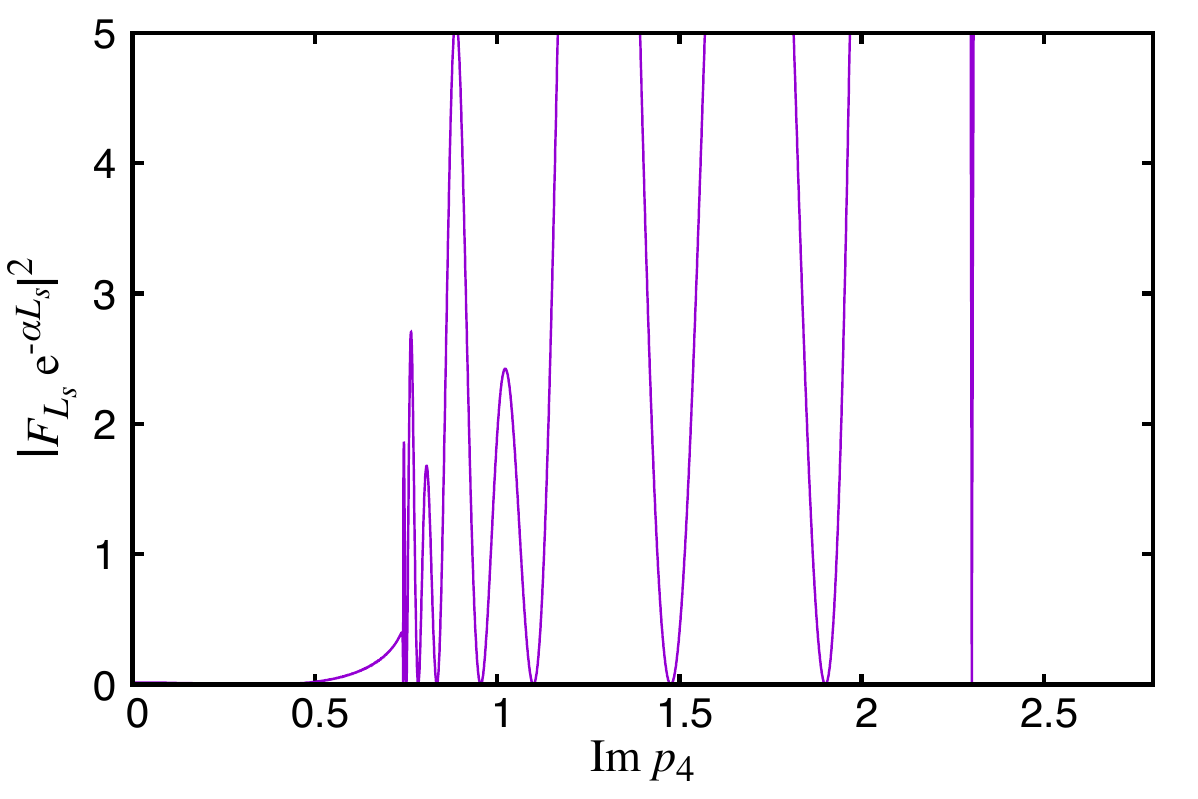}}}}
\subfigure{\mbox{\raisebox{1mm}{\includegraphics[width=99mm, bb=0 0 345 230]{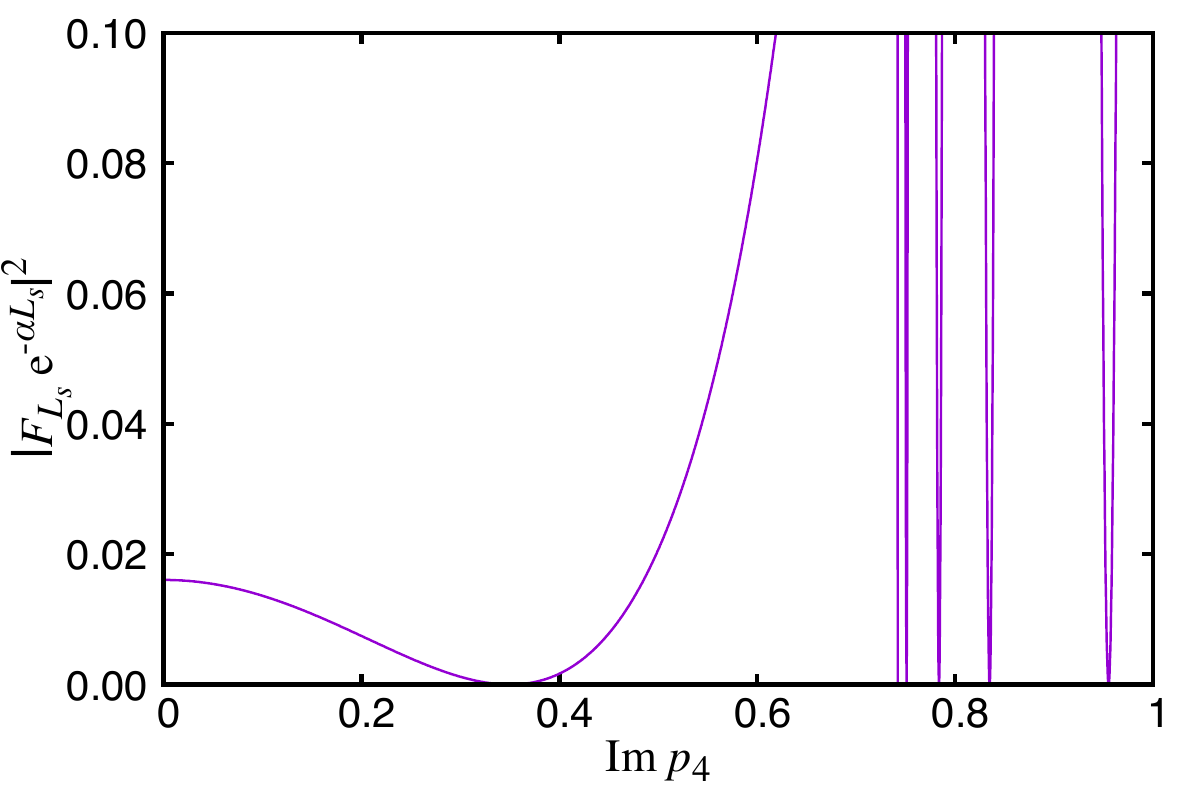}}}}
\subfigure{\mbox{\raisebox{1mm}{\includegraphics[width=99mm, bb=0 0 345 230]{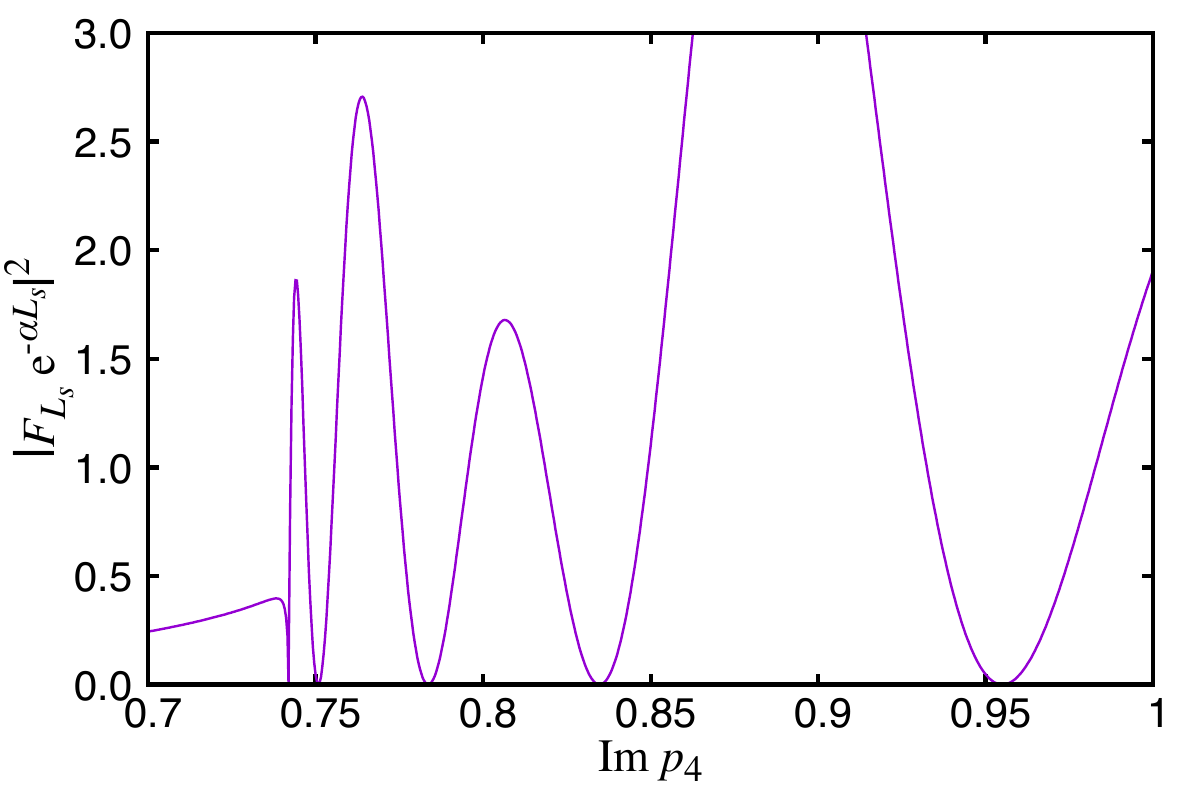}}}}
\caption{
$|F_{L_s}\e^{-\alpha L_s}|^2$ calculated at $L_s=8$, $M_5=0.9$, $m_f^{\rm pole}=0.35$,
$b-c=1$, $b+c=1$, $\vec p = 0$ and ${\rm Re}~p_4 = 0$ plotted as a function of
${\rm Im}~p_4$.
The lower two panels are magnifications of complicated parts in the top panel,
which accommodates all the zero points of $F_{L_s}$.
}
\label{fig:FLs}
\end{center}
\end{figure}

Besides this physical pole, we find that $F_{L_s}$ has other zero points,
which could give an unphysical contribution to four-dimensional physics.
Figure~\ref{fig:FLs} shows $|F_{L_s}\e^{-\alpha L_s}|^2$ calculated with
Shamir domain-wall fermions at $L_s=8$, $M_5=0.9$, $m_f^{\rm pole}=0.35$,
$\vec p=0$ and ${\rm Re}~p_4=0$.
While the physical pole is seen at ${\rm Im}~p_4=m_f^{\rm pole}=0.35$,
there are nine other zero points of $F_{L_s}$.
Two of them are trivially identified as the points satisfying $\cosh\alpha=1$ or
$\cosh\alpha=-1$, which correspond to ${\rm Im}~p_4 \simeq 2.30$ and 0.74 in
the plot, respectively.
Between these two zero points, the seven other zero points are found.
All of the zero points in this parameter choice are located on the imaginary
axis of $p_4$.

In \cite{Liang:2013eoa,Sufian:2016cft}, one of the trivial zero points
satisfying $\cosh\alpha=1$ was regarded as the unphysical pole
of domain-wall fermions.
However, the vanishing of $F_{L_s}$ at $\cosh\alpha=\pm1$ does not
mean the presence of unphysical poles at these points because
the numerator of the quark propagator \eqref{eq:qprop_4d} also
vanishes at these points and one can verify the limit
$\lim_{\alpha\rightarrow0,\img\pi} S_F^{4d}(p)$ is still finite.
In fact, the original $L_s\times L_s$ matrix $Q^\pm = (G^\pm)^{-1}$ is
still regular, ${\rm det}~Q^\pm\neq0$, even at these points.
This confusion may originate from the fact that the functional form of the inverse
matrix \eqref{eq:G_comp} is invalid for some special cases,
$\cosh\alpha=\pm1$ or $W=0$, and the inverse matrix $G_{s,t}^\pm$
in these special cases has another functional form as given in
Appendix~\ref{sec:qprop_special}.

The quark propagator at each of the remaining seven zero points
between these special zero points has a real singularity.
These zero points may give a significant lattice artifact when the
calculation is done at a large value of $m_f^{\rm pole}$.
We regard these zero points as the unphysical poles.
Note that these unphysical poles are located in the region $-1<\cosh\alpha<1$, in which
$\alpha$ is pure imaginary and any terms in \eqref{eq:FLs} are not suppressed
at large $L_s$, showing some oscillatory behavior with varying ${\rm Im}~p_4$.
As the extent of the fifth direction $L_s$ increases, the number of
these oscillations also increases, leading to the presence of more unphysical poles.
In our analysis, there are always $L_s-1$ unphysical poles.
In the limit of infinite $L_s$, an unphysical branch cut appears instead of a
series of unphysical poles as shown in Appendix~\ref{sec:infLs}.

In the following section, we discuss the fundamental properties of these unphysical
poles by showing the energy-momentum dispersion relation at various
input parameters.

\section{Dispersion relations}
\label{sec:disp_rel}

As mentioned in the previous section, the quark propagator \eqref{eq:qprop_4d}
at finite $L_s$ has $L_s-1$ unphysical poles as well as the physical pole.
In this section, we discuss the properties of the unphysical poles and the best
choice of the M\"obius parameters to suppress them, by analyzing the
dispersion relation for these poles in free field theory.
While it is likely familiar to the reader, for completeness we point out
that although lattice calculations are performed in Euclidean space with
$\tilde p^2 = \sum_\mu\sin^2p_\mu > 0$ it is the location of poles at negative
values of $p^2_4$ that determines the physical energies
$E(\vec p\, )=$ Im~$p_4(\vec p\, )$ of the quark states in free
field theory.  These poles and corresponding energies determine the
exponential fall-off of the quark propagators at large Euclidean time separations.
The dispersion relation for fermions on the lattice deviate from those for
continuum fermions with $O(a^2)$ error.
While dispersion relations for improved overlap fermions using the Brillouin kernel
were investigated \cite{Cho:2015ffa,Durr:2017wfi},
we concentrate on the dispersion relations for unimproved M\"obius domain-wall
fermions to investigate another source of cutoff effects due to unphysical poles.

\begin{figure}[t]
\begin{center}
\includegraphics[width=105mm, bb=0 0 345 230]{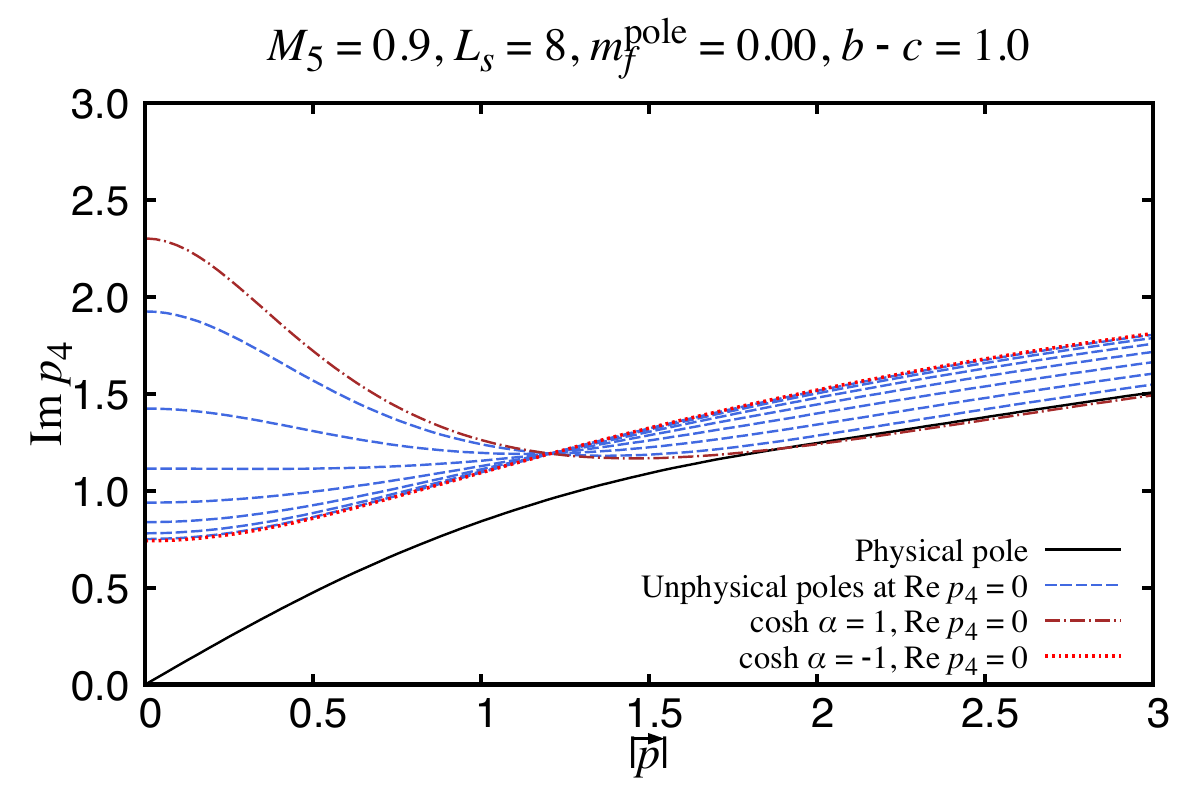}
\caption{
Dispersion relation for the domain-wall fermion at
$M_5=0.9, L_s=8, m_f^{\rm pole}=0, b+c=1, b-c=1$ and spatial momentum
$\vec p = ({|\vec p\,|\over\sqrt{3}}, {|\vec p\,|\over\sqrt{3}}, {|\vec p\,|\over\sqrt{3}})$.
}
\label{fig:UPP_M50.9}
\end{center}
\end{figure}

Figure~\ref{fig:UPP_M50.9} shows the dispersion relation for the domain-wall
fermion at $M_5 = 0.9$, $L_s=8$, $m_f^{\rm pole}=0$, $b=1$ and $c=0$.
The spatial momentum is chosen in the diagonal direction,
$\vec p = ({|\vec p\,|\over\sqrt{3}}, {|\vec p\,|\over\sqrt{3}}, {|\vec p\,|\over\sqrt{3}})$.
In the figure, one physical (solid curve) and seven unphysical poles (dashed curves)
are seen on the imaginary axis of $p_4$ at any spatial momentum.

As discussed in the previous section, these unphysical poles are confined to
the region between two curves, $\cosh\alpha=1$ (dashed-dotted curve) and
$\cosh\alpha=-1$ (dotted curve).
The boundaries $\cosh\alpha=\pm1$ are analytically given by
\begin{align}
\cos p_4|_{\cosh\alpha=1} &= \frac{\sum_{i=1}^3\sin^2p_i+B^2+1}{2B},
\label{eq:sol_cosha_p1}
\\*
\cos p_4|_{\cosh\alpha=-1}
&= \frac{4+4(b-c)B+(b-c)^2(\sum_{i=1}^3\sin^2p_i+B^2+1)}{4(b-c)+2(b-c)^2B},
\label{eq:sol_cosha_m1}
\\
B &= 4 -M_5 - \sum_{i=1}^3\cos p_i.
\end{align}
The solution of $\cosh\alpha=1$ depends only on $M_5$ and $p_i$,
implying that either of the corresponding lower or upper bounds on the unphysical
pole locations depends only on $M_5$.
On the other hand, the solution of $\cosh\alpha=-1$ depends also on $b-c$ and
therefore the other bound on the unphysical pole masses could be controlled by
varying $b-c$.
Since $b+c$ is not related to the region of unphysical pole energies and is usually
tuned to minimize the residual mass, we fix $b+c=1$ and do not vary it in this work.

\begin{figure}[t]
\begin{center}
\includegraphics[width=105mm, bb=0 0 345 230]{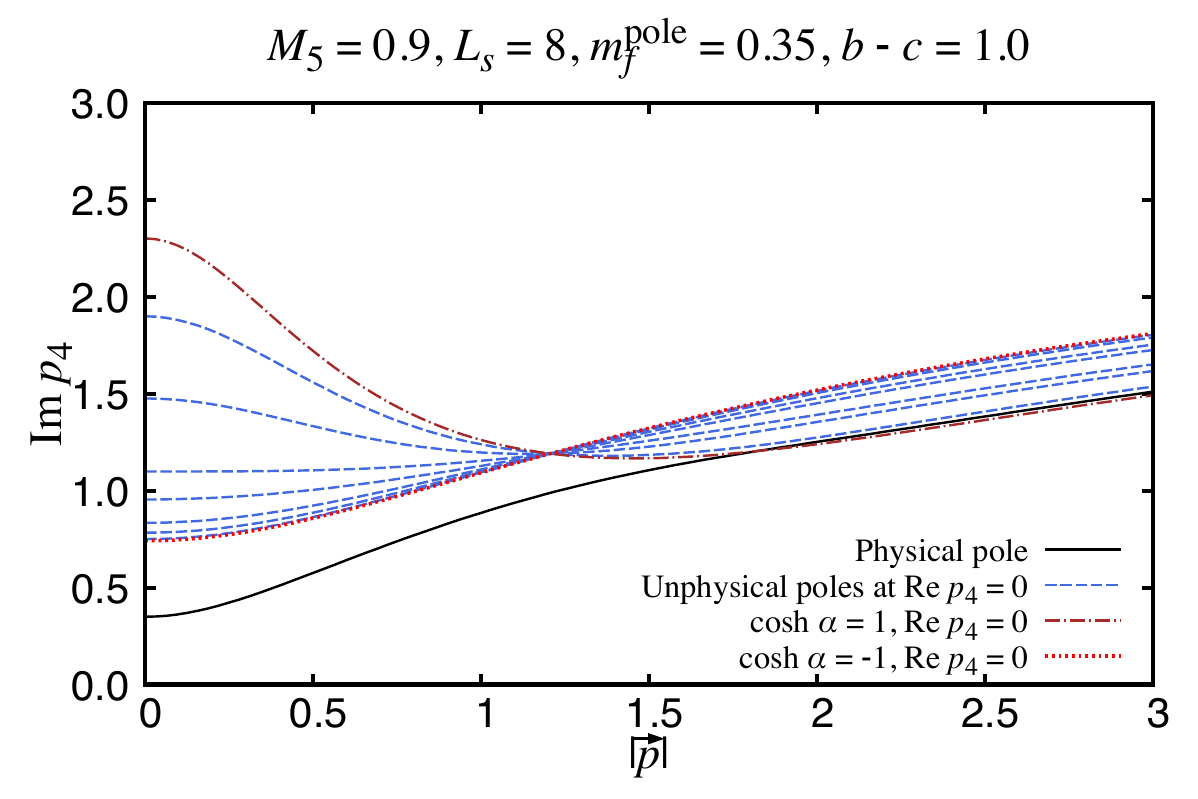}
\caption{
Same as Figure~\ref{fig:UPP_M50.9} but at $m_f^{\rm pole}=0.35$.
}
\label{fig:UPP_M50.9_mf}
\end{center}
\begin{center}
\includegraphics[width=105mm, bb=0 0 345 230]{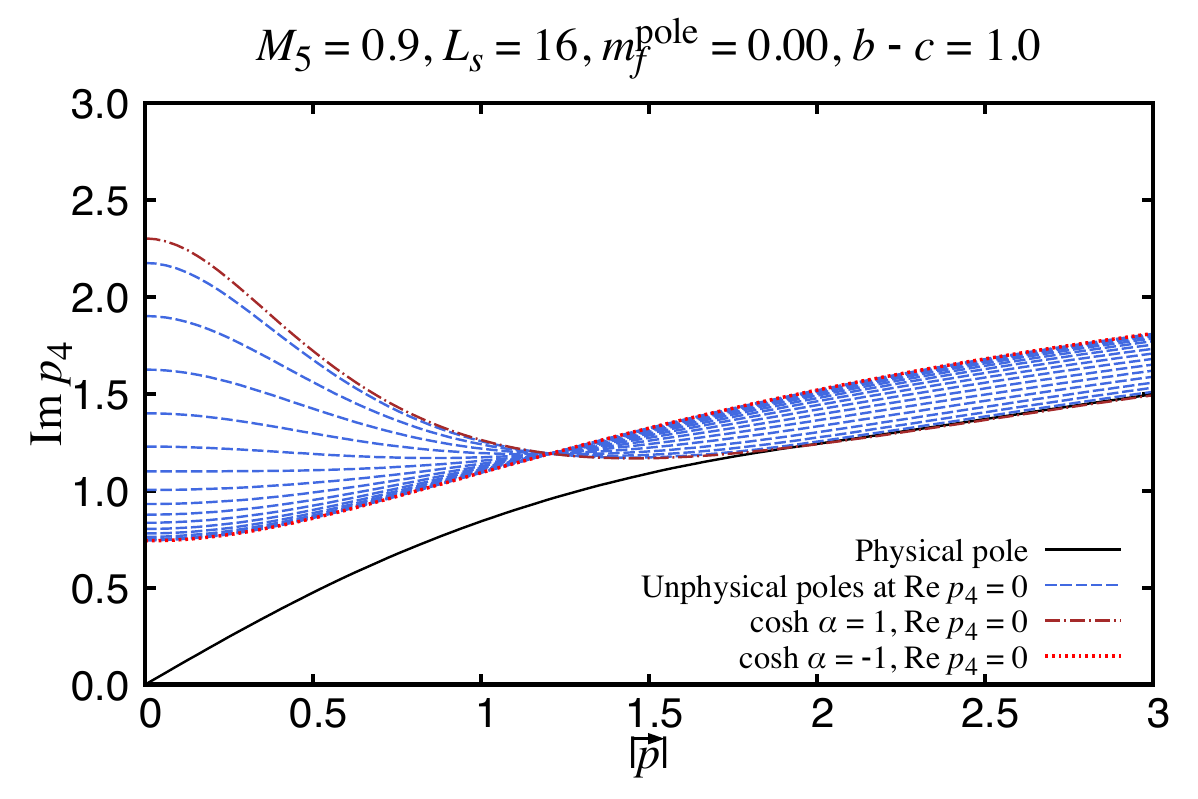}
\caption{
Same as Figure~\ref{fig:UPP_M50.9} but at $L_s=16$.
}
\label{fig:UPP_M50.9_Ls}
\end{center}
\end{figure}

Before varying $M_5$ and $b-c$, which play a key role to change the
region of unphysical pole energies, we briefly present the results of varying
the other parameters $m_f^{\rm pole}$ and $L_s$.
In Figure~\ref{fig:UPP_M50.9_mf}, we show the dispersion relation in a
massive case at $m_f^{\rm pole}=0.35$.
While the physical pole mass has certainly moved to 0.35, the unphysical
poles remain at close to their earlier locations when $m_f^{\rm pole}=0$.
In fact, the boundaries \eqref{eq:sol_cosha_p1} \eqref{eq:sol_cosha_m1}
of the unphysical poles are independent of $m_f^{\rm pole}$.
Therefore, as the physical pole mass $m_f^{\rm pole}$ increases and
approaches the unphysical pole masses, the dominance of the physical pole
would be lost.
A similar observation was shown in \cite{Liu:2003kp,Christ:2004gc}, which
investigated the physical and unphysical modes of the hermitian version of
the five-dimensional operator, the Dirac operator multiplied by $\gamma_5$
and the reflection operator.

Figure~\ref{fig:UPP_M50.9_Ls} shows the dispersion relation at $L_s=16$
and with the same values of the other parameters as those in
Figure~\ref{fig:UPP_M50.9}.
As described in the previous section, $F_{L_s}$ oscillates in the region
$-1<\cosh\alpha<1$ with varying the momentum variable and its frequency
is proportional to $L_s$.
Thus, the number of unphysical poles has been increased to 15.

\begin{figure}[t]
\begin{center}
\includegraphics[width=105mm, bb=0 0 345 230]{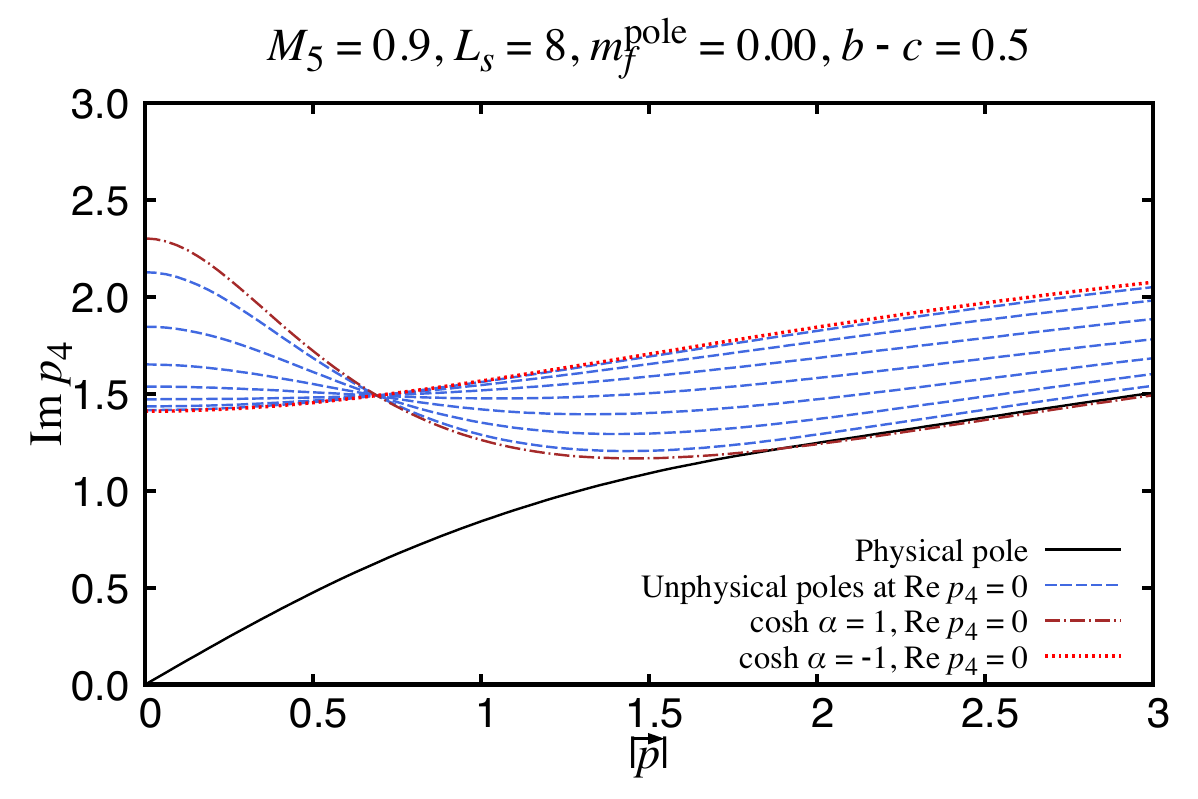}
\caption{
Same as Figure~\ref{fig:UPP_M50.9} but at $b-c=0.5$.
}
\label{fig:UPP_M50.9_c0.5}
\end{center}
\begin{center}
\includegraphics[width=105mm, bb=0 0 345 230]{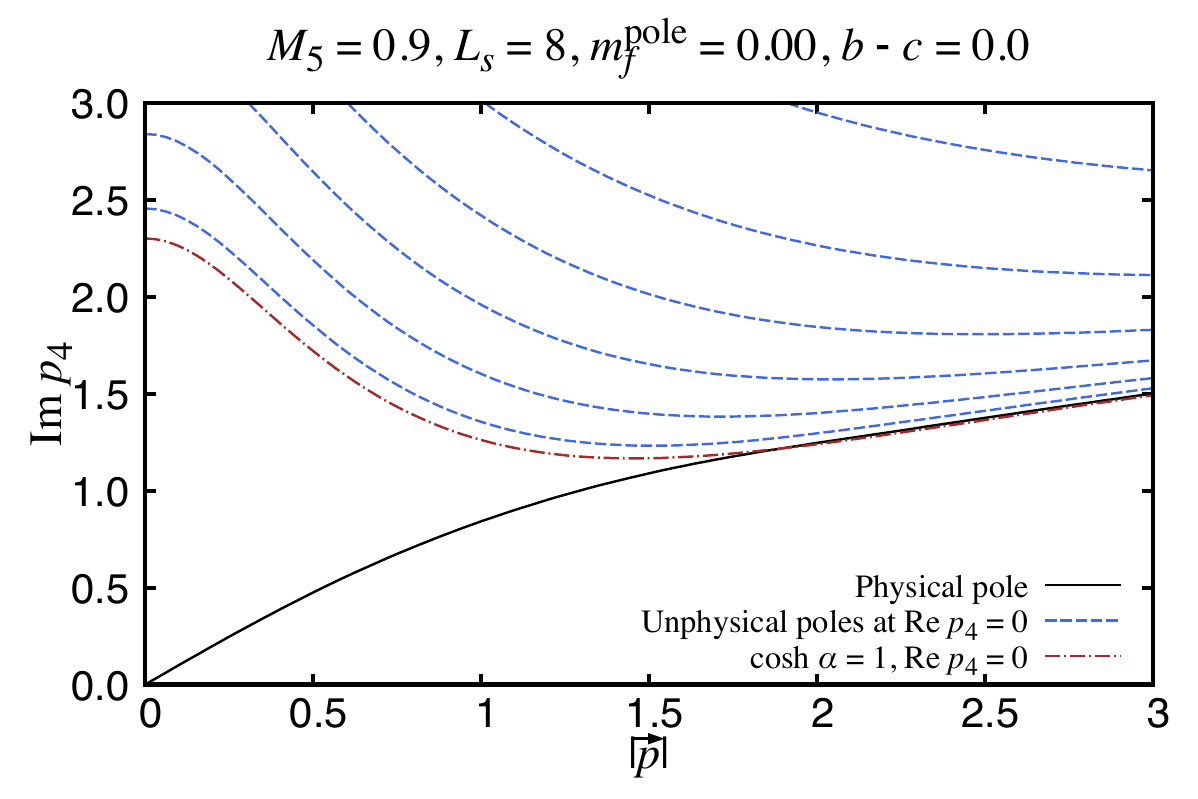}
\caption{
Same as Figure~\ref{fig:UPP_M50.9} but at $b-c=0$.
}
\label{fig:UPP_M50.9_c1.0}
\end{center}
\end{figure}

Now we show the results at smaller values of $b-c$.
Figure~\ref{fig:UPP_M50.9_c0.5} shows the result at $b-c=0.5$.
The values of ${\rm Im}~p_4$ on the curve $\cosh\alpha=-1$ are larger than
those for the Shamir type $b-c=1$ and the lower bound on the unphysical pole
masses has been increased to $\sim1.41$.
This fact implies that the contribution of unphysical poles at long distances would
be suppressed more rapidly.
Figure~\ref{fig:UPP_M50.9_c1.0} shows the result at $b-c=0$, where
${\rm Im}~p_4$ with $\cosh\alpha=-1$ is infinitely large as
\eqref{eq:sol_cosha_m1} indicates.
Thus, small values of $b-c$ make the unphysical modes heavy and realize
a small coupling between unphysical poles and the four-dimensional physics.

So far, we have discussed in the case of $M_5=0.9<1$, which has
the most simple structure of unphysical poles.
The case $M_5=1$ gives a similar dispersion relation with a slight modification
that ${\rm Im}~p_4$ with $\cosh\alpha=1$ diverges at $\vec p=0$ as described
by \eqref{eq:sol_cosha_p1}.

\begin{figure}[t]
\begin{center}
\includegraphics[width=105mm, bb=0 0 345 230]{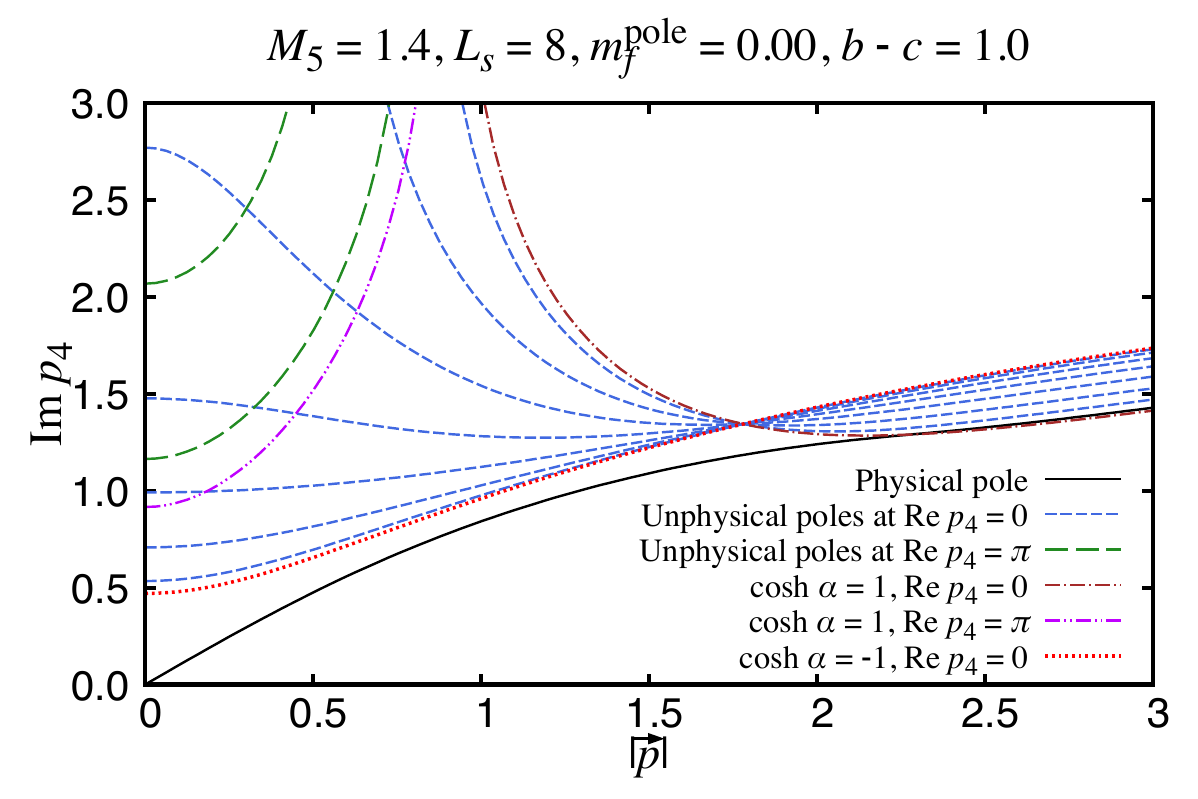}
\caption{
Same as Figure~\ref{fig:UPP_M50.9} but at $M_5=1.4$.
}
\label{fig:UPP_M51.4}
\end{center}
\end{figure}

In the case of $M_5>1$, $\alpha$ could be pure imaginary at
${\rm Re}~p_4=\pi$ as well as at ${\rm Re}~p_4=0$ and therefore
some of the unphysical poles are located at ${\rm Re}~p_4=\pi$.
Figure~\ref{fig:UPP_M51.4} shows the result at $M_5=1.4$.
The curve of $\cosh\alpha=1$ (dashed-dotted curve) on the imaginary axis
blows up at $|\vec p|\simeq0.90$, below which the solution of $\cosh\alpha=1$
(dashed double-dotted curve) is located at ${\rm Re}~p_4=\pi$.
Thus, there are some unphysical poles at ${\rm Re}~p_4=\pi$
(coarse-dashed curves) at small spatial momenta.
As suggested in \cite{Liang:2013eoa,Sufian:2016cft}, this kind of unphysical
pole may cause unphysical oscillation because the contribution of an unphysical
pole at $p_4=p_4^{\rm pole}$ to the quark propagator for the time direction has a
term $\sim\e^{\img p_4^{\rm pole}x_4}$, which is oscillatory unless
${\rm Re}~p_4^{\rm pole}=0$.

\begin{figure}[t]
\begin{center}
\includegraphics[width=105mm, bb=0 0 345 230]{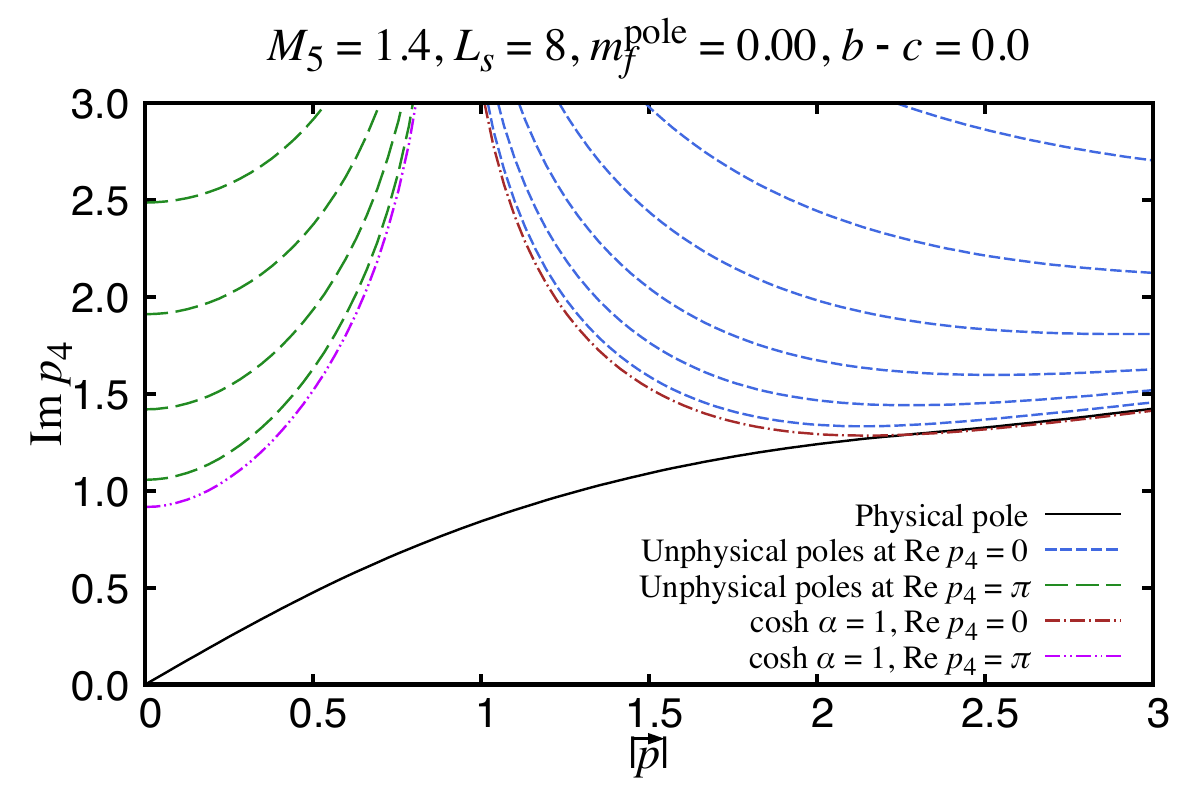}
\caption{
Same as Figure~\ref{fig:UPP_M51.4} but at $b-c=0$.
}
\label{fig:UPP_M51.4_c0.0}
\end{center}
\end{figure}

In Figure~\ref{fig:UPP_M51.4}, the lower bound on the unphysical
poles masses at ${\rm Re}~p_4=0$ is smaller than that at ${\rm Re}~p_4=\pi$,
implying that the unphysical contributions from the former poles are
more significant than those from the latter poles.
Figure~\ref{fig:UPP_M51.4_c0.0} shows the result at $b-c=0$.
Although the unphysical poles on the imaginary axis of $p_4$ at small spatial
momenta have certainly disappeared by taking $b-c=0$, all unphysical poles
have entered the region of ${\rm Re}~p_4=\pi$, whose lower limit
\eqref{eq:sol_cosha_p1} can be increased only by changing $M_5$.

We close this section with some comments on the choice of domain-wall parameters.
As we have seen, taking small $b-c$ plays a crucial role in reducing the
contribution of unphysical poles by increasing the lower bound on their masses.
In fact, the oscillatory behavior of domain-wall fermions \cite{Dudek:2006ej}, which is
supposed to be due to the unphysical poles, has never been observed in the case of
$b-c=0$, while that for $b-c=1$ is quite visible at large values of $M_5$.

The lower bound on the unphysical pole masses is determined by the solution of
$\cosh\alpha=-1$ \eqref{eq:sol_cosha_m1} if $b-c$ satisfies
\begin{align}
b-c &> \frac{2(1-M_5)}{1-(1-M_5)^2}\ \ \ \ {\rm for}\ M_5<1,
\notag\\
b-c &> \frac{2(M_5-1)}{1+(M_5-1)^2}\ \ \ \ {\rm for}\ M_5>1.
\label{ineq:bc_threshold}
\end{align}
At small values of $b-c$ that do not satisfy the inequality \eqref{ineq:bc_threshold},
the lower bound is determined by the solution of $\cosh\alpha=1$
\eqref{eq:sol_cosha_p1}, which is independent of $b-c$ and depends only on $M_5$.
This fact provides two prospects.
One is that taking extremely small $b-c$ compared to the threshold in
\eqref{ineq:bc_threshold} may not have a strong advantage.
The other is that $M_5$ may also need to be chosen appropriately
to suppress the unphysical contribution.
Obviously, the choice $M_5=1$ is optimal in free field theory.
In the mean field approximation \cite{Aoki:1997xg,Liang:2013eoa},
the optimal choice is modified to $M_5 = 4-3u_0$
with $u_0$ being the averaged link variable.

It is also important to take into account the violation of
chiral symmetry of the light quarks due to finite $L_s$.
The parameters $b+c$ and $M_5$ are usually tuned to minimize the residual
mass, while the dependence on $b-c$ has not been investigated a lot.
Note that small values of $b-c$, which are desired to reduce the
contribution of the unphysical poles, set a large upper limit on the
eigenvalues of the M\"obius kernel, potentially resulting in an inappropriate
approximation to the sign function.
Thus, the parameters $b,c$ and $M_5$ need to be carefully tuned in
non-perturbative studies so that both the residual mass and the contribution
of unphysical poles are safely small.

\section{Conclusion}
\label{sec:conclusion}

This study is dedicated to the exploration of a new way to precisely calculate
heavy-quark physics using M\"obius domain-wall fermions.
Our strategy is to treat the charm quark with the same regularization
as the lighter quarks without applying any effective theory or changing
any discretization parameters to achieve an appropriate GIM cancellation.
We have concentrated on a serious discretization error for heavy
quarks which originates from the unphysical poles of domain-wall fermions
by analyzing the energy-momentum dispersion relation.

As we have shown, the quark propagator constructed from domain-wall
fermions has $L_s-1$ unphysical poles and their energies are strongly
dependent on the difference of the M\"obius parameters $b-c$ as well
as on the domain-wall height $M_5$.
The lower bound on the unphysical pole masses in the case of $b-c=1$
is usually smaller than the lattice cutoff and quite comparable to
the charm-quark mass on lattices at currently available lattice spacings.
We demonstrated that this lower bound can be increased by taking
$b-c$ smaller.

One concern is that small $b-c$ could increase the residual mass because
the upper limit on the eigenvalues of the M\"obius kernel increases as
$b-c$ decreases, potentially spoiling the accuracy of the approximated
sign function.
We therefore need to tune the parameters taking account of the residual
breaking of chiral symmetry as well as of the impact of unphysical poles.
A non-perturbative study to explore the best choice of these parameters
is on-going.

\begin{acknowledgments}
I thank the members of the RBC and UKQCD collaborations and Katsumasa
Nakayama for useful discussions and comments.
I also express my gratitude to Norman~Christ and Raza~Sufian for
careful reading of the manuscript.
This work is supported in part by the US DOE grant \#DE-SC0011941.
\end{acknowledgments}

\appendix

\section{Unphysical branch cut in infinite $L_s$}
\label{sec:infLs}

In this appendix, we demonstrate what happens in the limit of infinite $L_s$.
First of all, it is important to note that the four-dimensional quark propagator
for infinite $L_s$ is supposed to be independent of $b+c$ because $b+c$ is
associated only with $L_s$-dependence, while \eqref{eq:qprop_4d_infLs} does
apparently contain $b+c$.
We can show \eqref{eq:qprop_4d_infLs} is independent of $b+c$ as follows.
Since the limit \eqref{eq:qprop_4d_infLs} is valid only if ${\rm Re}~\alpha>0$ on the
real axis of $p_4$, we can identify
\begin{equation}
\sinh\alpha = \frac{(b+c)\sqrt{FG}}{WD_-^\dag D_-},
\end{equation}
where we define
\begin{equation}
F = \frac{\tilde p^2+{\cal M}^2}{2},\ \ 
G = \frac{(b-c)^2\tilde p^2+(2+(b-c){\cal M})^2}{2}.
\end{equation}
In the case of $W<0$,
$\sinh\alpha$ is negative for ${\rm Re}~\alpha>0$ because $\cosh\alpha$ is also
negative and $\alpha$ has an imaginary part $\img\pi$.
Inserting $\e^{\pm\alpha} = \cosh\alpha\pm\sinh\alpha$ into \eqref{eq:qprop_4d_infLs},
we obtain
\begin{equation}
\lim_{L_s\rightarrow\infty}S_F^{4d}
=-\frac{\img\slashchar{\tilde p}+m({\cal M}+(b-c)F-\sqrt{FG})}
{(1-m^2)({\cal M}+(b-c)F)+(1+m^2)\sqrt{FG}}.
\label{eq:qprop_4d_infLs2}
\end{equation}
Thus, the four-dimensional quark propagator in the limit of infinite $L_s$ is independent
of $b+c$.

Since there are square roots of $FG$ in \eqref{eq:qprop_4d_infLs2}, some ambiguity
could occur in the region where $FG$ is not positive definite.
This ambiguity could be interpreted as the presence of an unphysical branch cut.
In fact, $F$ and $G$ vanish at $p_4$ satisfying $\cosh\alpha=1$ \eqref{eq:sol_cosha_p1}
and $\cosh\alpha=-1$ \eqref{eq:sol_cosha_m1}, respectively, and the product $FG$ is
negative between these two points.
Thus, a series of unphysical poles at finite $L_s$ becomes an unphysical
branch cut in the limit of infinite $L_s$.
The quark propagator at finite $L_s$ does not contain such a branch cut because
the insertion of $\e^{\pm\alpha} = \cosh\alpha\pm\sinh\alpha$ to \eqref{eq:qprop_4d}
cancels the square roots.

The propagator \eqref{eq:qprop_4d_infLs2} can be derived also from the Dirac
operator of overlap fermions, which is defined by
\begin{equation}
D_{\rm ov} = \frac{1+m}{2}+\frac{1-m}{2}\gamma_5 \frac{H_M}{\sqrt{H_M^2}}.
\end{equation}
Here, we use the M\"obius kernel
\begin{equation}
H_M = \gamma_5\frac{(b+c)D_W}{2+(b-c)D_W}.
\end{equation}
The inverse matrix of $D_{\rm ov}$ is found to be
\begin{equation}
D_{\rm ov}^{-1}
= \frac{(1-m)(-\img\slashchar{\tilde p} + {\cal M}+(b-c)F)+(1+m)\sqrt{FG}}
{(1-m^2)({\cal M}+(b-c)F) + (1+m^2)\sqrt{FG}},
\end{equation}
and obeys
\begin{equation}
\lim_{L_s\rightarrow\infty}S_F^{4d} = {D_{\rm ov}^{-1}-1\over1-m}.
\end{equation}
Therefore, overlap fermions have the same unphysical effects as domain-wall fermions.

\section{Propagator in some special cases}
\label{sec:qprop_special}

In this paper, we wrote the explicit form of $G^\pm$, the inverse of the matrix
\begin{equation}
Q^\pm = \bigg({b+c\over D_-^\dag D_-}\bigg)^2\tilde p^2+W^\mp W^\pm.
\end{equation}
The components of $Q^\pm$ are given by
\begin{align}
Q^\pm_{s,t} &= \left[\bigg({b+c\over D_-^\dag D_-}\bigg)^2\tilde p^2+W^2+1\right]\delta_{s,t}
-W(\delta_{s+1,t}+\delta_{s-1,t})
\notag\\&\hspace{4mm}
+mW(\delta_{s,L_s-1}\delta_{t,0}+\delta_{s,0}\delta_{t,L_s-1})
-(1-m^2)\times\Bigg\{
\begin{array}{l}
\delta_{s,0}\delta_{t,0}\ \ (+)
\\
\delta_{s,L_s-1}\delta_{t,L_s-1}\ \ (-)
\end{array}.
\end{align}

The inverse matrix $G^\pm=(Q^\pm)^{-1}$ satisfies the recurrence
relations
\begin{align}
\left[\bigg({b+c\over D_-^\dag D_-}\bigg)^2\tilde p^2+W^2+1\right]G^\pm_{s,t}
-W(G^\pm_{s+1,t}+G^\pm_{s-1,t}) = \delta_{s,t},
\label{eq:recurr_rel1}
\\
\left[\bigg({b+c\over D_-^\dag D_-}\bigg)^2\tilde p^2+W^2+1\right]G^\pm_{s,t}
-W(G^\pm_{s,t+1}+G^\pm_{s,t+1}) = \delta_{s,t},
\label{eq:recurr_rel2}
\end{align}
and the boundary conditions
\begin{align}
&WG^+_{-1,t}-(1-m^2)G_{0,t}^++mWG_{L_s-1,t}^+=0,
\label{eq:BC_init}
\\
&G_{L_s,t}^++mG_{0,t}^+=0,
\\
&mG_{L_s-1,t}^-+G_{-1,t}^-=0,
\\
&WG_{L_s,t}^--(1-m^2)G_{L_s-1,t}^-+mWG_{0,t}^-=0.
\label{eq:BC_last}
\end{align}

The solution for the usual case is already given in \eqref{eq:G_comp}.
In the following, we give $G^\pm_{s,t}$ and the corresponding four-dimensional
quark propagator $S_F^{4d}$ in the special cases,
$W=0$, $\cosh\alpha=1$ and $\cosh\alpha=-1$.

\subsection{$W=0$}

In this case, the matrices $Q^\pm$ are given by
\begin{align}
Q_{s,t}^+
&= \left[\bigg(\frac{b+c}{D^\dag_-D_-}\bigg)^2\tilde p^2+1\right]\delta_{s,t}
-(1-m^2)\delta_{s,0}\delta_{t,0},
\\
Q_{s,t}^-
&= \left[\bigg(\frac{b+c}{D^\dag_-D_-}\bigg)^2\tilde p^2+1\right]\delta_{s,t}
-(1-m^2)\delta_{s,L_s-1}\delta_{t,L_s-1}.
\end{align}
The corresponding inverse matrices are
\begin{align}
G^+_{s,t}
&=\left[\bigg(\frac{b+c}{D^\dag_-D_-}\bigg)^2\tilde p^2+1\right]^{-1}\delta_{s,t}(1-\delta_{s,0})
+\left[\bigg(\frac{b+c}{D^\dag_-D_-}\bigg)^2\tilde p^2+m^2\right]^{-1}\delta_{s,0}\delta_{t,0},
\\
G^-_{s,t}
&=\left[\bigg(\frac{b+c}{D^\dag_-D_-}\bigg)^2\tilde p^2+1\right]^{-1}\delta_{s,t}(1-\delta_{s,L_s-1})
+\left[\bigg(\frac{b+c}{D^\dag_-D_-}\bigg)^2\tilde p^2+m^2\right]^{-1}\delta_{s,L_s-1}\delta_{t,L_s-1},
\end{align}
which correspond to the four-dimensional quark propagator
\begin{equation}
S_F^{4d} = \frac{-\frac{b+c}{D_-^\dag D_-}\img\slashchar{\tilde p}+m}
{\big(\frac{b+c}{D_-^\dag D_-}\big)^2\tilde p^2+m^2}.
\end{equation}

\subsection{$\cosh\alpha=1$}

In this case,
the recurrence relations \eqref{eq:recurr_rel1} and \eqref{eq:recurr_rel2} are
\begin{align}
2WG_{s,t}^\pm-W(G_{s+1,t}^\pm+G_{s-1,t}^\pm)=\delta_{s,t},\ \ 
2WG_{s,t}^\pm-W(G_{s,t+1}^\pm+G_{s,t-1}^\pm)=\delta_{s,t},
\end{align}
whose solution is formally given by
\begin{align}
G_{s,t}^\pm &= -\frac{|s-t|}{2W} + C^{(1)}_{2,\pm} st + C^{(1)}_{s,\pm} s + C^{(1)}_{t,\pm} t + C^{(1)}_{0,\pm},
\end{align}
The boundary conditions \eqref{eq:BC_init}--\eqref{eq:BC_last} determine
the coefficients,
\begin{align}
C^{(1)}_{2,+}&=C^{(1)}_{2,-}
= -\frac{1}{WF^{(1)}_{L_s}}(1-m^2)(1-W),
\\
C^{(1)}_{s,+}&=C^{(1)}_{t,+}
= \frac{1}{2WF_{L_s}^{(1)}}(1-m^2)((1-W)L_s+W),
\\
C^{(1)}_{s,-}&=C^{(1)}_{t,-} = \frac{1}{2WF_{L_s}^{(1)}}(1-m^2)((1-W)L_s+W-2),
\\
C^{(1)}_{0,+} &= -\frac{L_s}{F_{L_s}^{(1)}},
\\
C^{(1)}_{0,-} &= -\frac{1}{WF_{L_s}^{(1)}}(WL_s-(1-m^2)(L_s-1)),
\\
F_{L_s}^{(1)} &= (1-m^2)(1-W)L_s-W(1+m)^2.
\end{align}
The corresponding four-dimensional propagator is given by
\begin{align}
S_F^{4d} &= \frac{b+c}{D_-^\dag D_-}\frac{L_s}{F_{L_s}^{(1)}}\img\slashchar{\tilde p}
-\frac{L_sm(1-W)+(1+m)W}{F_{L_s}^{(1)}}
\notag\\
&\xrightarrow{\ L_s\rightarrow\infty\ }
\frac{\frac{b+c}{D_-^\dag D_-}\img\slashchar{\tilde p}-m(1-W)}{(1-m^2)(1-W)}
=-\frac{\img\slashchar{\tilde p}-m{\cal M}}{(1-m^2){\cal M}}.
\end{align}
The inverse matrix $G^\pm_{s,t}$ and the four-dimensional quark propagator
$S_F^{4d}$ derived in this subsection can be reproduced also from the limit
$\alpha\rightarrow0$ of $G^\pm_{s,t}$ and $S_F^{4d}$ in the standard case
\eqref{eq:G_comp} \eqref{eq:qprop_4d} \eqref{eq:qprop_4d_infLs}.

\subsection{$\cosh\alpha=-1$}

The recurrence relations \eqref{eq:recurr_rel1} and \eqref{eq:recurr_rel2} in this case are
\begin{align}
-2WG_{s,t}^\pm-W(G_{s+1,t}^\pm+G_{s-1,t}^\pm)=\delta_{s,t},\ \ 
-2WG_{s,t}^\pm-W(G_{s,t+1}^\pm+G_{s,t-1}^\pm)=\delta_{s,t},
\end{align}
whose solution is formally given by
\begin{align}
G_{s,t}^\pm = 
\left(
\frac{|s-t|}{2W} + C^{(-1)}_{2,\pm} st + C^{(-1)}_{s,\pm} s + C^{(-1)}_{t,\pm} t + C^{(-1)}_{0,\pm}
\right)(-1)^{s-t},
\end{align}
In the case of even $L_s$, the coefficients are
\begin{align}
C^{(-1)}_{2,+}&=C^{(-1)}_{2,-}
= \frac{1}{WF^{(-1)}_{L_s}}(1-m^2)(1+W),
\\
C^{(-1)}_{s,+}&=C^{(-1)}_{t,+}
= -\frac{1}{2WF_{L_s}^{(-1)}}(1-m^2)((1+W)L_s-W),
\\
C^{(-1)}_{s,-}&=C^{(-1)}_{t,-} = -\frac{1}{2WF_{L_s}^{(-1)}}(1-m^2)((1+W)L_s-W-2),
\\
C^{(-1)}_{0,+} &= -\frac{L_s}{F_{L_s}^{(-1)}},
\\
C^{(-1)}_{0,-} &= -\frac{1}{WF_{L_s}^{(-1)}}(WL_s+(1-m^2)(L_s-1)),
\\
F_{L_s}^{(-1)} &= (1-m^2)(1+W)L_s+W(1+m)^2.
\end{align}
The corresponding four-dimensional propagator is given by
\begin{align}
S_F^{4d} &= \frac{b+c}{D_-^\dag D_-}\frac{L_s}{F_{L_s}^{(-1)}}\img\slashchar{\tilde p}
-\frac{L_sm(1+W)-(1+m)W}{F_{L_s}^{(-1)}}
\notag\\
&\xrightarrow{\ L_s\rightarrow\infty\ }
\frac{\frac{b+c}{D_-^\dag D_-}\img\slashchar{\tilde p}-m(1+W)}{(1-m^2)(1+W)}
= -\frac{2\img\slashchar{\tilde p}+m(2{\cal M}+(b-c)(\tilde p^2+{\cal M}^2))}
{(1-m^2)(2{\cal M}+(b-c)(\tilde p^2+{\cal M}^2)}.
\end{align}
$G^\pm_{s,t}$ and $S_F^{4d}$ derived in this subsection can be reproduced also
from the limit $\alpha\rightarrow\img\pi$ of $G^\pm_{s,t}$ and $S_F^{4d}$ in the
standard case \eqref{eq:G_comp} \eqref{eq:qprop_4d} \eqref{eq:qprop_4d_infLs}.

\bibliography{DWfreeUPP}

\providecommand{\href}[2]{#2}\begingroup\raggedright\begin{thebibliography}{10}

\bibitem{Glashow:1970gm}
S.~L. Glashow, J.~Iliopoulos, and L.~Maiani, ``{Weak Interactions with
  Lepton-Hadron Symmetry},''
\href{http://dx.doi.org/10.1103/PhysRevD.2.1285}{{\em Phys. Rev.} {\bfseries
  D2} (1970) 1285--1292}.

\bibitem{Brower:2004xi}
R.~C. Brower, H.~Neff, and K.~Orginos, ``{Mobius fermions: Improved domain wall
  chiral fermions},''
  \href{http://dx.doi.org/10.1016/j.nuclphysbps.2004.11.180}{{\em Nucl. Phys.
  Proc. Suppl.} {\bfseries 140} (2005) 686--688},
  \href{http://arxiv.org/abs/hep-lat/0409118}{{\ttfamily arXiv:hep-lat/0409118
  [hep-lat]}}.
[,686(2004)].

\bibitem{Brower:2012vk}
R.~C. Brower, H.~Neff, and K.~Orginos, ``{The M\'obius Domain Wall Fermion
  Algorithm},''
\href{http://arxiv.org/abs/1206.5214}{{\ttfamily arXiv:1206.5214 [hep-lat]}}.

\bibitem{Kaplan:1992bt}
D.~B. Kaplan, ``{A Method for simulating chiral fermions on the lattice},''
  \href{http://dx.doi.org/10.1016/0370-2693(92)91112-M}{{\em Phys. Lett.}
  {\bfseries B288} (1992) 342--347},
\href{http://arxiv.org/abs/hep-lat/9206013}{{\ttfamily arXiv:hep-lat/9206013
  [hep-lat]}}.

\bibitem{Shamir:1993zy}
Y.~Shamir, ``{Chiral fermions from lattice boundaries},''
  \href{http://dx.doi.org/10.1016/0550-3213(93)90162-I}{{\em Nucl. Phys.}
  {\bfseries B406} (1993) 90--106},
\href{http://arxiv.org/abs/hep-lat/9303005}{{\ttfamily arXiv:hep-lat/9303005
  [hep-lat]}}.

\bibitem{Chen:2014hva}
{\bfseries TWQCD} Collaboration, W.-P. Chen, Y.-C. Chen, T.-W. Chiu, H.-Y.
  Chou, T.-S. Guu, and T.-H. Hsieh, ``{Decay Constants of Pseudoscalar
  $D$-mesons in Lattice QCD with Domain-Wall Fermion},''
  \href{http://dx.doi.org/10.1016/j.physletb.2014.07.025}{{\em Phys. Lett.}
  {\bfseries B736} (2014) 231--236},
\href{http://arxiv.org/abs/1404.3648}{{\ttfamily arXiv:1404.3648 [hep-lat]}}.

\bibitem{Boyle:2017jwu}
P.~A. Boyle, L.~Del~Debbio, A.~Juttner, A.~Khamseh, F.~Sanfilippo, and J.~T.
  Tsang, ``{The decay constants ${\mathbf{f_D}}$ and ${\mathbf{f_{D_{s}}}}$ in
  the continuum limit of ${\mathbf{N_f=2+1}}$ domain wall lattice QCD},''
\href{http://arxiv.org/abs/1701.02644}{{\ttfamily arXiv:1701.02644 [hep-lat]}}.

\bibitem{Yang:2014sea}
Y.-B. Yang {\em et~al.}, ``{Charm and strange quark masses and $f_{D_s}$ from
  overlap fermions},'' \href{http://dx.doi.org/10.1103/PhysRevD.92.034517}{{\em
  Phys. Rev.} {\bfseries D92} no.~3, (2015) 034517},
\href{http://arxiv.org/abs/1410.3343}{{\ttfamily arXiv:1410.3343 [hep-lat]}}.

\bibitem{Bai:2014cva}
Z.~Bai, N.~H. Christ, T.~Izubuchi, C.~T. Sachrajda, A.~Soni, and J.~Yu,
  ``{$K_L-K_S$ Mass Difference from Lattice QCD},''
  \href{http://dx.doi.org/10.1103/PhysRevLett.113.112003}{{\em Phys. Rev.
  Lett.} {\bfseries 113} (2014) 112003},
\href{http://arxiv.org/abs/1406.0916}{{\ttfamily arXiv:1406.0916 [hep-lat]}}.

\bibitem{Christ:2014qwa}
{\bfseries RBC, UKQCD} Collaboration, N.~Christ, T.~Izubuchi, C.~T. Sachrajda,
  A.~Soni, and J.~Yu, ``{Calculating the $K_L-K_S$ mass difference and
  $\epsilon_K$ to sub-percent accuracy},'' {\em PoS} {\bfseries LATTICE2013}
  (2014) 397,
\href{http://arxiv.org/abs/1402.2577}{{\ttfamily arXiv:1402.2577 [hep-lat]}}.

\bibitem{Bai:2016gzv}
Z.~Bai, ``{Long distance part of $\epsilon_K$ from lattice QCD},'' {\em PoS}
  {\bfseries LATTICE2016} (2017) 309,
\href{http://arxiv.org/abs/1611.06601}{{\ttfamily arXiv:1611.06601 [hep-lat]}}.

\bibitem{Christ:2015aha}
{\bfseries RBC, UKQCD} Collaboration, N.~H. Christ, X.~Feng, A.~Portelli, and
  C.~T. Sachrajda, ``{Prospects for a lattice computation of rare kaon decay
  amplitudes: $K\to\pi\ell^+\ell^-$ decays},''
  \href{http://dx.doi.org/10.1103/PhysRevD.92.094512}{{\em Phys. Rev.}
  {\bfseries D92} no.~9, (2015) 094512},
\href{http://arxiv.org/abs/1507.03094}{{\ttfamily arXiv:1507.03094 [hep-lat]}}.

\bibitem{Christ:2016mmq}
N.~H. Christ, X.~Feng, A.~Juttner, A.~Lawson, A.~Portelli, and C.~T. Sachrajda,
  ``{First exploratory calculation of the long-distance contributions to the
  rare kaon decays $K\to\pi\ell^+\ell^-$},''
  \href{http://dx.doi.org/10.1103/PhysRevD.94.114516}{{\em Phys. Rev.}
  {\bfseries D94} no.~11, (2016) 114516},
\href{http://arxiv.org/abs/1608.07585}{{\ttfamily arXiv:1608.07585 [hep-lat]}}.

\bibitem{Christ:2016eae}
{\bfseries RBC, UKQCD} Collaboration, N.~H. Christ, X.~Feng, A.~Portelli, and
  C.~T. Sachrajda, ``{Prospects for a lattice computation of rare kaon decay
  amplitudes II $K\to\pi\nu\bar{\nu}$ decays},''
  \href{http://dx.doi.org/10.1103/PhysRevD.93.114517}{{\em Phys. Rev.}
  {\bfseries D93} no.~11, (2016) 114517},
\href{http://arxiv.org/abs/1605.04442}{{\ttfamily arXiv:1605.04442 [hep-lat]}}.

\bibitem{Bai:2017fkh}
Z.~Bai, N.~H. Christ, X.~Feng, A.~Lawson, A.~Portelli, and C.~T. Sachrajda,
  ``{Exploratory lattice QCD study of the rare kaon decay
  $K^+\to\pi^+\nu\bar\nu$},''
\href{http://arxiv.org/abs/1701.02858}{{\ttfamily arXiv:1701.02858 [hep-lat]}}.

\bibitem{Liu:2003kp}
G.-f. Liu, {\em {Quark eigenmodes and lattice QCD}}.
\newblock PhD thesis, Columbia U., 2003.
\newblock
\url{http://wwwlib.umi.com/dissertations/fullcit?p3104827}.
\newblock

\bibitem{Christ:2004gc}
N.~H. Christ and G.~Liu, ``{Massive domain wall fermions},''
  \href{http://dx.doi.org/10.1016/S0920-5632(03)02553-2}{{\em Nucl. Phys. Proc.
  Suppl.} {\bfseries 129} (2004) 272--274}.
[,272(2004)].

\bibitem{Dudek:2006ej}
J.~J. Dudek, R.~G. Edwards, and D.~G. Richards, ``{Radiative transitions in
  charmonium from lattice QCD},''
  \href{http://dx.doi.org/10.1103/PhysRevD.73.074507}{{\em Phys. Rev.}
  {\bfseries D73} (2006) 074507},
\href{http://arxiv.org/abs/hep-ph/0601137}{{\ttfamily arXiv:hep-ph/0601137
  [hep-ph]}}.

\bibitem{Syritsyn:2007mp}
S.~Syritsyn and J.~W. Negele, ``{Oscillatory terms in the domain wall transfer
  matrix},'' {\em PoS} {\bfseries LAT2007} (2007) 078,
\href{http://arxiv.org/abs/0710.0425}{{\ttfamily arXiv:0710.0425 [hep-lat]}}.

\bibitem{Liang:2013eoa}
J.~Liang, Y.~Chen, M.~Gong, L.-C. Gui, K.-F. Liu, Z.~Liu, and Y.-B. Yang,
  ``{Oscillatory behavior of the domain wall fermions revisited},''
  \href{http://dx.doi.org/10.1103/PhysRevD.89.094507}{{\em Phys. Rev.}
  {\bfseries D89} no.~9, (2014) 094507},
\href{http://arxiv.org/abs/1310.3532}{{\ttfamily arXiv:1310.3532 [hep-lat]}}.

\bibitem{Sufian:2016cft}
R.~S. Sufian, M.~J. Glatzmaier, and Y.-B. Yang, ``{Unphysical Poles of Domain
  Wall Fermions at finite $L_s$},''
\href{http://arxiv.org/abs/1603.01591}{{\ttfamily arXiv:1603.01591 [hep-lat]}}.

\bibitem{Borici:1999zw}
A.~Borici, ``{Truncated overlap fermions},''
  \href{http://dx.doi.org/10.1016/S0920-5632(00)91802-4}{{\em Nucl. Phys. Proc.
  Suppl.} {\bfseries 83} (2000) 771--773},
\href{http://arxiv.org/abs/hep-lat/9909057}{{\ttfamily arXiv:hep-lat/9909057
  [hep-lat]}}.

\bibitem{Blum:2014tka}
{\bfseries RBC, UKQCD} Collaboration, T.~Blum {\em et~al.}, ``{Domain wall QCD
  with physical quark masses},''
  \href{http://dx.doi.org/10.1103/PhysRevD.93.074505}{{\em Phys. Rev.}
  {\bfseries D93} no.~7, (2016) 074505},
\href{http://arxiv.org/abs/1411.7017}{{\ttfamily arXiv:1411.7017 [hep-lat]}}.

\bibitem{Hernandez:1998et}
P.~Hernandez, K.~Jansen, and M.~Luscher, ``{Locality properties of Neuberger's
  lattice Dirac operator},''
  \href{http://dx.doi.org/10.1016/S0550-3213(99)00213-8}{{\em Nucl. Phys.}
  {\bfseries B552} (1999) 363--378},
\href{http://arxiv.org/abs/hep-lat/9808010}{{\ttfamily arXiv:hep-lat/9808010
  [hep-lat]}}.

\bibitem{Kikukawa:2000kd}
Y.~Kikukawa and Y.~Nakayama, ``{Gauge anomaly cancellations in SU(2)(L) x
  U(1)(Y) electroweak theory on the lattice},''
  \href{http://dx.doi.org/10.1016/S0550-3213(00)00714-8}{{\em Nucl. Phys.}
  {\bfseries B597} (2001) 519--536},
\href{http://arxiv.org/abs/hep-lat/0005015}{{\ttfamily arXiv:hep-lat/0005015
  [hep-lat]}}.

\bibitem{Aoki:2002vt}
Y.~Aoki {\em et~al.}, ``{Domain wall fermions with improved gauge actions},''
  \href{http://dx.doi.org/10.1103/PhysRevD.69.074504}{{\em Phys. Rev.}
  {\bfseries D69} (2004) 074504},
\href{http://arxiv.org/abs/hep-lat/0211023}{{\ttfamily arXiv:hep-lat/0211023
  [hep-lat]}}.

\bibitem{Antonio:2008zz}
{\bfseries RBC, UKQCD} Collaboration, D.~J. Antonio {\em et~al.},
  ``{Localization and chiral symmetry in three flavor domain wall QCD},''
  \href{http://dx.doi.org/10.1103/PhysRevD.77.014509}{{\em Phys. Rev.}
  {\bfseries D77} (2008) 014509},
\href{http://arxiv.org/abs/0705.2340}{{\ttfamily arXiv:0705.2340 [hep-lat]}}.

\bibitem{Hagler:2007xi}
{\bfseries LHPC} Collaboration, P.~Hagler {\em et~al.}, ``{Nucleon Generalized
  Parton Distributions from Full Lattice QCD},''
  \href{http://dx.doi.org/10.1103/PhysRevD.77.094502}{{\em Phys. Rev.}
  {\bfseries D77} (2008) 094502},
\href{http://arxiv.org/abs/0705.4295}{{\ttfamily arXiv:0705.4295 [hep-lat]}}.

\bibitem{WalkerLoud:2008bp}
A.~Walker-Loud {\em et~al.}, ``{Light hadron spectroscopy using domain wall
  valence quarks on an Asqtad sea},''
  \href{http://dx.doi.org/10.1103/PhysRevD.79.054502}{{\em Phys. Rev.}
  {\bfseries D79} (2009) 054502},
\href{http://arxiv.org/abs/0806.4549}{{\ttfamily arXiv:0806.4549 [hep-lat]}}.

\bibitem{Cho:2015ffa}
Y.-G. Cho, S.~Hashimoto, A.~Jüttner, T.~Kaneko, M.~Marinkovic, J.-I. Noaki,
  and J.~T. Tsang, ``{Improved lattice fermion action for heavy quarks},''
  \href{http://dx.doi.org/10.1007/JHEP05(2015)072}{{\em JHEP} {\bfseries 05}
  (2015) 072},
\href{http://arxiv.org/abs/1504.01630}{{\ttfamily arXiv:1504.01630 [hep-lat]}}.

\bibitem{Boyle:2016imm}
P.~Boyle, A.~Juttner, M.~K. Marinkovic, F.~Sanfilippo, M.~Spraggs, and J.~T.
  Tsang, ``{An exploratory study of heavy domain wall fermions on the
  lattice},'' \href{http://dx.doi.org/10.1007/JHEP04(2016)037}{{\em JHEP}
  {\bfseries 04} (2016) 037},
\href{http://arxiv.org/abs/1602.04118}{{\ttfamily arXiv:1602.04118 [hep-lat]}}.

\bibitem{Aoki:1997xg}
S.~Aoki and Y.~Taniguchi, ``{One loop calculation in lattice QCD with domain
  wall quarks},'' \href{http://dx.doi.org/10.1103/PhysRevD.59.054510}{{\em
  Phys. Rev.} {\bfseries D59} (1999) 054510},
\href{http://arxiv.org/abs/hep-lat/9711004}{{\ttfamily arXiv:hep-lat/9711004
  [hep-lat]}}.

\end{thebibliography}\endgroup
\end{document}